\documentstyle[12pt]{article}
\input epsf
\begin{document}
\baselineskip16pt
\thispagestyle{empty}
\date{}
\title{Bianchi V inflation \\ in the Brans-Dicke theory?}
\author{J.L. Cervantes-Cota\thanks{e-mail:jorge@nuclear.inin.mx} \\
Departamento de F\'{\i}sica, \\
Instituto Nacional de Investigaciones Nucleares (ININ) \\
P.O. Box 18-1027, M\'exico D.F. 11801, M\'exico} 
\maketitle
\newpage
\begin{abstract}
It is shown some exact solutions in the Brans-Dicke (BD) 
theory for a Bianchi V metric having the property of 
inflationary expansion, graceful exit, and asymptotic 
evolution to a Friedmann-Robertson-Walker (FRW) open 
model.  It is remarkable that an inflationary behaviour 
can occur, even without a cosmological potential or 
constant.   However, the horizon and flatness problems 
cannot be solve within the standard BD theory because the 
inflationary period is severely restricted by the value of the 
BD parameter $\omega$.  

\end{abstract}

\newpage

\section{INTRODUCTION \label{intro}} 
It is well known that the Universe is homogeneous and isotropic on very
large scales.  Supporting this assertion are different measurements as the
isotropy measured in the Cosmic Microwave Background Radiation (CMBR) by   
the Cosmic Background Explorer (COBE) satellite \cite{Cobe94}; isotropy in 
x-ray backgrounds (e.g., quasars at high red shift); and in number counts
in faint radio sources, cf. Ref. \cite{KoTu90}.  All these measurements give 
e\-vi\-dence for the cosmological principle, which consistently 
should be valid on very large scales of our Universe.  However, 
on smaller scales there are a variety of inhomogeneities 
and anisotropies that have been produced by grow of 
matter perturbations, at least since the time of last scattering 
until now.  Accor\-dingly, many authors have tried to model 
the early Universe with non-FRW metrics allowing general initial 
conditions to explain, after its evolution, small and large scale
properties of our Universe.  Thus, the chaotic cosmo\-lo\-gy programme 
\cite{Mi67} tried, unsuccessfully, to isotropize anisotropic cosmologies 
in  order to understand, within general footings, 
the present large scale isotropy.  Collins and Hawking \cite{CoHa73} proved, 
within General Relativity (GR),  that if the dominant energy condition and
positive presssure criterion are satisfied, the universe can approach 
isotropy only if it is one of the types I, V, VII$_{0}$, and VII$_{h}$.  However, at finite times other anisotropic models can be bounded from 
above \cite{BaSo86} and even they are consistent with COBE measurements \cite{Ba95}, that is, anisotropic cosmologies do not need 
inflationary scenarios to explain the observed 
$\frac{\Delta T}{T} \, \approx \, 10^{-5}$ in the CMBR.  However, a 
causal origin of the perturbations is not provided by anisotropic 
cosmologies alone, and inflationary cosmologies does provide 
such an explanation, up to some fine tuning.  This is one reason 
why inflation is still the most interesting proposal to the standard 
model of cosmology.

The question of the entry to an inflationary phase, now known as the cosmic 
no hair conjecture \cite{nh-conj}, as well as the question of its exit 
(graceful exit problem) are two key aspects of inflation that have been 
intensively discussed.  No hair theorems have been proved in the 
context of GR within anisotropic
and some inhomogeneous cosmologies \cite{nh-models}.  The  entry to inflation
has been also discussed in some other contexts, including tilt matter 
models \cite{AnMaRoRy91}, and  some anisotropic models in scalar tensor 
theories \cite{anist}.  In general, one has realized that for 
some set of initial conditions there exists an inflationary attractor for 
the evolution.  However, the exit of inflation is not necessarily guaranteed
when a cosmological constant ($\Lambda$) is present.  Instead of $\Lambda$ 
one has been considering a cosmological function $V(\phi)$ that plays the role
of a constant during some time $\tau=N H^{-1}$, where $N$ is the number of 
e-foldings of exponential expansion and $H \approx$const. the Hubble 
parameter.  Adding a properly chosen constant\footnote{So fine tuned to avoid 
having a cosmological constant nowadays ($n$) bigger than 
$8 \pi G \rho_{n} = 3 H_{n}^{2} \sim 10^{-83} {\rm GeV}{}^{2}$.} to $V(\phi)$ 
one gets  after inflation that $V(\phi_f) \approx 0$, then the Universe 
preheats \cite{KoLiSt95}, reheats, and experiences a metamorphosis to a 
FRW dynamics filled with a particle content characterized by the decay of 
the $\phi-$particles.  The introduction of a constant in $V(\phi)$, to 
allow inflation to occur and to graceful exit, is well motived but it 
is put by hand.  In this sense, it would be desirable to have a model 
in which both features, the entry and exit of inflation, arise 
{\it naturally}, as a consequence of the exact, integrating 
dynamics.  In GR, however,  one needs to introduce a 
cosmological constant (or function) to generate inflation.  In 
scalar-tensor theories this term appears naturally, but to obtain  
successful inflation it implies the fine tuning of parameters that not
always are in accord with the particle physics models behind.  
   
Recently, effort has been put on the analysis of FRW models in the 
Brans-Dicke (BD) theory  \cite{BrDi61}, which is simplest gravity theory 
beyond GR.   In Refs. \cite{Ko95-6,HoWa98} a qualitative analysis of the 
general evolution and asymptotic behaviour is studied, as well as the 
conditions for inflation even without a cosmological function or 
constant, see also Ref. \cite{Le95}.   Motived by these analyses we study 
in the present work Bianchi V anisotropic solutions in the BD 
theory, because it is one of the simplest step  further in 
complexity and permits the analysis of the anisotropic properties of 
models; in particular the isotropization of some solutions has been studied 
in Refs. \cite{ChCe95,MiWa95}.   We found that Bianchi V models can have 
the properties of inflation (without a potential or a cosmological 
constant), graceful exit of the inflationary stage, and evolution to an 
effective FRW open Universe, like the one we observe nowadays.  The 
range of values for the coupling constant $\omega$ is, however, very 
different from what one would desire, i.e. $\omega > 500$ \cite{Wi93}.

In the next section we show the analytic solutions and, thereafter, 
in section \ref{solana} we analyse their behaviour.  In section \ref{fluc} 
some comments are made on fluctuations of the BD field in these models. 
Finally, section \ref{con} is devoted to conclusions.

\section{BRANS-DICKE BIANCHI V SO\-LU\-TIONS \label{anis}}

The BD theory \cite{BrDi61} has the following Einstein-Hilbert action, 
with signature (+,-,-,-):
\begin{equation}
\label{bd}
{\cal L}= \left[ \phi \, R  + 
\frac{\omega}{\phi} \, \phi_{| \mu} \phi^{| \mu} + 16 \pi L_{M} \right] 
\frac{\sqrt{-g}}{16 \pi}
\end{equation}
where $R$ is the Ricci scalar, $\phi$ the 
scalar field, the symbol $|$ partial derivative, $\omega$ the coupling 
constant of the theory, and $g$ the determinant of the metric 
tensor.  After varying this equation one derives the BD field 
equations
\begin{eqnarray}
\label{gr}
R_{\mu \nu} -\frac{1}{2} R g_{\mu \nu}  \, & = & \,
-  \frac{8 \pi}{\phi} \, T_{\mu \nu} - \frac{\omega}{\phi^{2}}  
 \left[ \phi_{| \mu} \phi_{| \nu}   -
 \frac{1}{2} \phi_{| \lambda} \phi^{| \lambda} \,\ g_{\mu \nu} \right]
 \nonumber\\ &&
- \frac{1}{\phi} 
\left[ \phi_{| \mu || \nu} -
\phi^{| \lambda}_{ \,\  \,\ || \lambda} \,\ g_{\mu \nu} \right] 
\end{eqnarray}
and  
\begin{equation}
\label{hig}
 \phi^{| \lambda }{}_{|| \lambda} \,\ = \,\ 
 \frac{8 \pi}{3+2\omega} \, T \,\ ,
\end{equation}
where the symbol $||$  stands for the covariant
derivative and $T$ is the trace of the energy-momentum tensor, 
$\, T_{\mu \nu }$.  The continuity equation (energy-momentum 
conservation law) reads
\begin{equation}
\label{eqct}
\, \, T^{\,\ \nu}_{\mu \,\ \,\ || \nu} \,\ = \,\ 0 \,\ .
\end{equation}

We consider homogeneous, aniso\-tro\-pic Bianchi-type models to study the 
dynamics, and particularly a Bianchi type V spacetime symmetry in a 
synchronous coordinate frame given by the line element \cite{Ma79}:
\begin{equation}
\label{mb5}
ds^{2}= dt^{2} - a_{1}^{2} \, dx^{2}  - a_{2}^{2} \, e^{-2x} dy^{2} - 
       a_{3}^{2} \, e^{-2x}dz^{2} \,\ .
\end{equation}
Following, we shall use scaled variables to find and analyse the solutions 
in a simple way: the scaled field $\psi \equiv \phi a^{3(1-\nu)}$, a new 
cosmic time parameter 
$d\eta \equiv a^{-3\nu} dt$, $()^\prime\equiv \frac{d}{d\eta}$, 
the `volume' $a^{3}\equiv a_{1}a_{2}a_{3}$, and the Hubble parameters 
$H_{i}\equiv {a_{i}}^\prime /a_{i}$ corresponding to the scale factors 
$a_{i}=a_{i}(\eta)$ for $i=1,2,3$.  We assume comoving coordinates and 
a barotropic equation of state for the perfect fluid 
present ($T_{\mu \nu}$), $p=\nu \rho$, with $\nu$ 
a constant. Using these definitions and the above given metric, one obtains 
the cosmological equations from Eqs. (\ref{gr}) and (\ref{hig}):
\begin{equation}
\label{a123}
(\psi H_i)^\prime -  \psi a^{6 \nu} C_{iV}  \, = \, 
\frac{8 \pi}{3+2\omega}  [1 + (1 - \nu) \omega] \, \rho  a^{3(1+\nu)}
 ~~~ {\rm for} ~~~ i=1,2,3. 
\end{equation}

\begin{eqnarray}
\label{h123} 
&&H_{1}H_{2} + H_{1}H_{3} + H_{2}H_{3} + 
[1+(1-\nu)\omega] \, \left(H_{1}+H_{2}+H_{3}\right) 
\frac{\psi^\prime}{\psi}
\nonumber \\[2pt]
&&- (1-\nu)[1+\omega(1-\nu)/2] (H_{1}+H_{2}+H_{3})^{2} 
-\frac{\omega}{2} \left( \frac{\psi^\prime}{\psi}\right)^{2} 
-\frac{C_{V}}{2} a^{6 \nu}  \nonumber \\[2pt]
&& = \,\ 8 \pi \, \frac{\rho a^{3(1+\nu)}}{\psi} \,  \,\ ,
\end{eqnarray}

\begin{equation}
\label{psi} 
\psi^{\prime \prime} + (\nu-1) a^{6 \nu} C_{V} \, \psi =  
\frac{8 \pi}{3+2\omega}  [2(2-3\nu)+3(1-\nu)^{2} \omega ] \, 
\rho  a^{3(1+\nu)} \,\ ,
\end{equation}
and
\begin{equation}
\label{h12}
H_{2} + H_{3} \,\ = \,\ 2 H_{1}  \,\ ,
\end{equation}
where $C_{V}\equiv \Sigma_{i} C_{iV}$ with the partial the curvature terms 
equal to each other, i.e., $C_{iV}=2/a_{1}^{2}$ for $i=1,2,3$.  
Eq. (\ref{h12}) implies that $a_{2}$ and $a_{3}$ are inverse proportional 
functions, $a_{2}a_{3}= a_{1}^{2}$; note that 
$3H\equiv H_{1}+H_{2}+H_{3}=3H_{1}$, $H$ being the mean Hubble parameter.

Additionally, the continuity equation yields:
\begin{equation}
\label{rho} 
\rho a^{3(1+\nu)}={\rm const.}\equiv M_{\nu}
\end{equation}
$M_{\nu}$ being a dimensional constant depending on the fluid present. The 
vacuum case is attained when $M_{\nu}=0$.
  
The system of ordinary differential equations, Eqs.  
(\ref{a123}-\ref{h12}), can be decoupled to obtain an equation for 
$\psi$ alone, by means of the relation $a_1 = a$, which can be once 
integrated to get:
\begin{eqnarray}
\label{psi1}
&&\psi \, \psi^{\prime \prime} -  \frac{2}{3(1-\nu)} {\psi^{\prime}}^{2} -
\frac{2(1-3\nu)}{3(1-\nu)} \, [m_{\nu} (1-3\nu) \eta + \eta_{o}] \, 
\psi^{\prime} + [2+(1-\nu)(1+3\nu)\omega]m_{\nu} \, \psi + \nonumber \\
&&\frac{2(1-3\nu)}{3(1-\nu)} \, [2(2-3\nu) + 3 (1-\nu)^{2} \omega] \, 
[\frac{m_{\nu}}{2} (1-3\nu) \, \eta^{2} + \eta_{o} \, \eta + 
\eta_{1}^{2}] m_{\nu} \,\ = \,\ K \,\ ,
\end{eqnarray}
where $\eta_{o}$, $\eta_{1}$, and $K$ are integration constants; 
$m_{\nu}\equiv \frac{8 \pi M_{\nu}}{3+2 \omega}$.  Solutions
to Eq. (\ref{psi1}) are to be used to get the mean Hubble parameter through
the following equation
\begin{equation}
\label{psih}
H_{1} + H_{2} + H_{3} \, = \, \frac{1}{1-\nu}  \,  
\left[ \frac{\psi^{\prime}}{\psi}  - 
\frac{(1-3\nu) m_{\nu} \, \eta + \eta_{o}}{\psi} \right]   \, ,
\end{equation}
obtained by using Eqs. (\ref{a123}) and (\ref{psi}).  Additionally, each
Hubble rate is written as follows (similar to the Bianchi type I model 
Ref. \cite{RuFi75}):
\begin{equation} 
\label{hi}
H_i \, = \, {1 \over 3} \left( H_{1}+H_{2}+H_{3} \right) + 
{h_i \over \psi } = 
\frac{\psi^{\prime} - (1-3\nu) m_{\nu} \, \eta -  \eta_{o} + 
3 (1-\nu) h_i}{3 (1-\nu) \psi} \, ,
\end{equation}
where the $h_i$'s are constants that determine the 
anisotropic character of the solutions. If $h_i=0$ for $i=1, 2$, $3$ 
simultaneously, no anisotropy is present.  Furthermore, the Bianchi V model 
obeys the condition
\begin{equation}
\label{resth123}
 h_1 + h_2 + h_3 = 0  , \quad {\rm with} \quad h_1 = 0, 
\end{equation}
the last relation stems from the fact that $H_{1} = H$, see Eq. (\ref{h12}).

In order to analyse the anisotropic character of the solutions, we have 
constructed the following `constraint' equation  for 
the anisotropic shear, $\sigma$:
\begin{eqnarray}
&& \sigma ( \eta) \equiv ~ - (H_1 - H_2)^2 - (H_2 - H_3)^2 - (H_3 - H_1)^2  
\,\ = \,\ \nonumber \\[4pt]
&&{3 \over {2(1-\nu)}}\left( {\psi^{\prime \prime} \over \psi} \right)-
{1\over{(1-\nu)^2}}\left( {\psi^{\prime} \over \psi} \right )^2 -
{{(1-3 \nu)}\over{(1-\nu)^2}} \left( {{ (1-3 \nu)m_\nu\eta + \eta_o}
\over \psi} \right) \left( {\psi^{\prime} \over \psi} \right )
\nonumber \\[4pt]
&& +{{[2-3 \nu+ {3\over2}\omega(1-\nu)^2]}
\over {(1-\nu)^2}} \left({{ (1-3 \nu)m_\nu\eta + \eta_o}
\over \psi}\right)^2+{{3 [2+\omega(1-\nu)(1+3\nu)] m_\nu}
\over{2(1-\nu) \psi}} 
\, . \label{con2}  \end{eqnarray}
$\sigma = 0$ is a
necessary condition to obtain a FRW cosmology since it implies 
$H_1 = H_2 = H_3$, cf. Ref. \cite{ChCeNu91,ChCe95}.  If the sum of 
the squared differences of the Hubble expansion rates 
tends to zero, it would mean that the anisotropic scale factors tend to 
a single function of time which is, certainly, the scale factor of the open 
Friedmann model.  

The anisotropic shear becomes, using Eq. (\ref{resth123}),
\begin{equation}
\label{sig}
\sigma(\eta) = - \frac{6 h_{2}^{2}}{\psi^2}
\,\  ,
\end{equation}
or the dimensionless shear parameter\cite{WaEl97}
\begin{equation}
\label{dlsig}
\frac{\sigma}{H^{2}} = - \frac{54 (1-\nu)^{2} h_{2}^{2}}
{[\psi^{\prime}-(1-3\nu) m_{\nu} \eta -\eta_{o}]^{2}}
\,\  .
\end{equation}
These equations admit solutions such that $\sigma \to 0$,  
$\sigma/H^{2} \to 0$ as $\eta \to \infty$  (or $t \to \infty$), that is, 
one has time asymptotic isotropization solutions, similar 
to the solutions found for the Bianchi V model in GR, 
see Ref. \cite{BaSo86}.  In fact, one does not need to impose an 
asymptotic, infinity condition, but just that $\eta \gg \eta_{*}$, 
where $\eta_{*}$ is yet some arbitrary value to warrant 
that $\sigma$ ($\sigma/H^{2}$) can be bounded from above.  

The above equations are valid for any value $\nu$ of the 
equation of state, but the solutions are different for 
$\nu \neq 1/3$ and $\nu = 1/3$.  Therefore, we treat both 
cases separately.
 
\subsection {Solutions with $ \nu \neq \frac{1}{3}$}

The explicit solutions are following given.  A particular solution of Eq.  
(\ref{psi1}) is \cite{ChCe95}:
\begin{equation}
\label{psisol}
\psi \,= \, A_{V} \, \eta^2 + B_{V} \, \eta + C_{V}  \, ,
\end{equation}
with the constants 
\begin{eqnarray}
\label{constab5}
A_{{}_V} & \,=\, & - {{(1-3\nu)^2}\over{(1+3\nu)}} \, m_\nu ~ ,\nonumber \\
B_{{}_V} & \,=\, & -2 \left({{1-3\nu}\over{1+3\nu}}\right) \, \eta_0 ~ , 
\end{eqnarray}
being the same as for the isotropic, $k=\pm 1$ cases\cite{ChCe95}, 
but now 
\begin{equation}
\label{rest5}
 m_\nu \, (1+3\nu) \, C_{{}_V} \,=\, -\, 
{(1+3\nu)^2 (h_1^2 + h_2^2 + h_3^2) \over 18 \nu + \omega (1+3\nu)^2} 
\,-\, \eta_0^2~~. \end{equation}
The constants $B_{{}_V}$ and $ C_{{}_V}$, being proportional 
to the $h_{i}'$s, encode information on the anisotropic 
shear, that is, on the nature of the ani\-so\-tro\-pic character 
of this Bianchi type model.  The constants $\eta_{1}$ and $K$ of 
Eq. (\ref{psi1}) are determined through Eq. (\ref{rest5}) but they 
have no further significance.

The Hubble expansion rates, given through Eqs. (\ref{hi}, 
\ref{psisol}, \ref{constab5}), are 
\begin{equation}
\label{h15}
 H_i(\eta) =  - {1 \over (1+3\nu)} 
{ {(1 - 3 \nu) m_\nu \eta + \eta_0 - (1+3\nu) h_{i}} \over 
( A_{{}_V} \eta^2 + B_{{}_V} \eta + C_{{}_V} )} \qquad {\rm for} \quad 
i=1,2,3 \end{equation}
which can be directly integrated to get the scale factors:
\begin{equation}
\label{a15}
a_1=\left[{-2(1+3\nu)}\over{[ 2+(1- \nu)(1+3\nu)\omega]}
{m_\nu}\right]^{1\over 2(1-3\nu)}
\left[ A_{{}_V}\eta^2+B_{{}_V}\eta+C_{{}_V} \right]^{1\over2( 1-3\nu)}
~~~, \end{equation}
\begin{equation}
\label{a25}
a_2=a_1\exp\left[{-2h_2 \over \sqrt{\Delta}}
{\rm arctanh} \left( {-2(1-3\nu)[(1-3\nu)m_\nu \eta + \eta_0] \over
(1+3\nu) \sqrt{\Delta}}\right) \right] , \,\ \Delta > 0 
\end{equation}
or
\begin{equation}
\label{a2n5}
a_2=a_1\exp\left[{ 2h_2 \over \sqrt{-\Delta}}
{\rm arctan} \left( {-2(1-3\nu)[(1-3\nu)m_\nu \eta + \eta_0] \over
(1+3\nu) \sqrt{-\Delta}}\right) \right] ,  \,\ \Delta < 0 
\end{equation}
and according to Eq. (\ref{h12}) one gets
\begin{equation}
\label{a2a3}
a_{3}=\frac{a_{1}^{2}}{a_{2}} ~~, \end{equation}
where the discriminant is
\begin{equation}
\label{dis5}
\Delta={B_{{}_V}^2-4A_{{}_V}C_{{}_V}} = {{-8 (1-3\nu)^2} \over 
{18 \nu+(1+3\nu)^2 \omega}}{h_2^2}  \,\ , 
\end{equation}  
being proportional to the shear, see Eq. (\ref{sig}).  Independent of the 
value $\Delta$ might have, $ h_1 + h_2 + h_3=0 $ is always true.  The 
type V model with $ \Delta \not = 0$ fulfills that 
$h_2=-h_3$ with $h_1 = 0$ to have truly anisotropic solutions.  But for 
$\Delta=0$, 
\begin{equation}
\label{aequal}
a_3=a_2=a_1 ~~, \end{equation}
since one has that $h_{1}=h_{2}=h_{3}=0$.  In this case, the scale factors 
are equal to each other, given by Eq. (\ref{a15}), up to a constant that can be 
scaled away. 

The BD field is
\begin{equation}
\label{phi5}
\phi = \frac{\psi}{a^{3(1-\nu)}} = 
\left[{[2+(1-\nu)(1+3\nu)\omega]m_\nu}\over{-2(1+3\nu)}
\right]^{3(1-\nu)\over 2(1-3\nu)}\left[A_{{}_V} \eta^2+B_{{}_V} 
\eta+C_{{}_V} \right]^{-{{1+3\nu}\over{2(1-3\nu)}}}~~.
\end{equation}

The above-presented solutions reduce to the previously known 
for the isotropic, FRW model with an open space $(k=-1)$, cf. 
Refs. \cite{DeOb70s,Le74,ChGu86}, when  $\Delta=0$, implying that
$h_1=h_2=h_3=0$, to have no shear, 
$\sigma = 0$.  The above solutions have no analogues in GR since 
the scalar field $\phi$ cannot be set to a constant.

It has been conjectured, for the Bianchi V dust ($\nu=0$) 
model, that it is almost impossible to find solutions because 
of the complexity of the equation for $\psi$,  which is presented 
in Ref. \cite{Lo84c} as a fourth-order differential 
equation.  Therefore, only particular solutions are expected to be 
found, if any.  In the present work it is shown an 
integrated equation for $\psi$ which is now of second-order, see 
Eq. (\ref{psi1}), and possesses at least one solution for $\nu\neq1/3$ 
given by Eq. (\ref{psisol}); other solutions are hoped to be found 
because of the relative simplicity of Eq. (\ref{psi1}).  The 
dust-solution in Ref. \cite{Lo84c} is written in other 
variables, but it is a special case of our above solution when one 
takes $B=\nu=0$.  Referring to this, one of the 
main problems of the physical interpretation in the field 
of exact solutions in cosmology has been that results are written 
in {\it scaled} variables and the physics behind them 
is partially hidden.  In our case, for the dust model ($\nu=0$) the scaled time
is the same as the original cosmic time, written 
in Eq. (\ref{mb5}).  For $\nu \neq 0$ effort should be made to 
interpret physical situations. 

Following we present the solutions for the radiation case.

\subsection {Solutions with $\nu=1/3$} 

In the radiation case the curvature terms 
in Eqs. (\ref{a123}, \ref{h123}, \ref{psi}) turn out to be 
especially simple ($a^{6 \nu} C_{iV}=2$, $a^{6 \nu} C_{V}=6$).  
Therefore,  Eq. (\ref{psi}) reduces to
\begin{equation}
\label{psi2}
\psi^{\prime \prime} -  4 \psi - \frac{16 \pi M_{ _{1/3}}}{3} = 0 \, , 
\end{equation}
that can be directly integrated to get\footnote{This solution 
is also valid for Eq. (\ref{psi1}) with $\nu=1/3$.} 
\begin{equation}
\label{psiradsol}
\psi = c_{1} e^{- 2 \eta} + c_{2} e^{2 \eta} - \frac{4 \pi M_{ _{1/3}}}{3}, 
\qquad c_{1}, \, c_{2} \quad {\rm are ~integration ~constants.}
\end{equation}

The general solution for the Hubble expansion rates, using Eqs. 
(\ref{psih}, \ref{hi}, \ref{psiradsol}), is given by
\begin{equation}
\label{hiradsol}
H_{i}(\eta) = 
\frac{-c_{1} e^{- 2 \eta} + c_{2} e^{2 \eta} - c_{3} + h_{i}}
{c_{1} e^{- 2 \eta} + c_{2} e^{2 \eta} - \frac{4 \pi M_{ _{1/3}}}{3}}
\qquad {\rm for} \quad i=1,2,3
\end{equation}
where $c_{3}\,(\equiv \eta_{o}/2)$ is an arbitrary integration 
constant and the constants $h_i$, accounting for the anisotropy 
of the models, are determined by: 
\begin{eqnarray}
\label{hiconsts}
&& h_1 = 0, \nonumber \\
&& h_2 = - h_3, \nonumber \\ 
&& h_3 = \mp \sqrt{12 \Delta - (3 + 2\omega) \, c_{3}^{2} } 
\end{eqnarray}
where 
$\Delta \equiv (2 \pi M_{ _{1/3}}/3)^{2} - c_{1} c_{2}$.  The
constants given in Eq. (\ref{hiconsts}) are consistent with 
Eqs. (\ref{hi}, \ref{resth123}, \ref{sig}).  For if 
$h_{1} = h_{2} = h_{3} = 0$, it implies that $\sigma = 0$.
 
{}From Eq. (\ref{hiradsol}) one obtains the scale factors
\begin{equation}
\label{airad}
a_{i} =\alpha _{io} \, \psi^{1/2} \,  
{\rm exp} \left[\frac{c_3 - h_{i}}{2 \sqrt{\Delta}} \, 
{\rm arctanh} \, 
\left(  \frac{1}{\sqrt{\Delta}}
(c_{2} e^{2\eta} - 2\pi M_{ _{1/3}}/3) \right) \right] ,  
\,\ \Delta > 0 \, ,
\end{equation}
\begin{equation}
\label{ainrad}
a_{i} = \alpha_{io} \, \psi^{1/2} \,  
{\rm exp} \left[ \frac{h_{i}-c_{3}}{2 \sqrt{-\Delta}} \, {\rm arctan} \, 
\left( \frac{1}{\sqrt{-\Delta}}
(c_{2} e^{2\eta} - 2\pi M_{ _{1/3}}/3) \right) \right] , \,\ \Delta < 0 \, ,
\end{equation}
or
\begin{equation}
\label{aid0rad}
a_{i} = \alpha_{io} \, \psi^{1/2} \,  
{\rm exp} \left[\frac{c_{3} -h_{i}}{2 c_{2} e^{2 \eta} -4 \pi M_{ _{1/3}}/3}  
\right] , \,\ \Delta = 0 \, ,
\end{equation}
where $\alpha_{io}$ are integration constants proportional to some initial 
size of the Universe.   Note that Eq. (\ref{a2a3}) is valid again.

The BD field ($\phi = \psi/ a^{2}$) is 
\begin{equation}
\label{phirad}
\phi = (\alpha_{1o} \alpha_{2o} \alpha_{3o})^{-2/3} 
{\rm exp} \left[-\frac{c_{3}}{\sqrt{\Delta}} \, {\rm arctanh} \, 
\left(\frac{1}{\sqrt{\Delta}}
(c_{2} e^{2\eta} - 2\pi M_{ _{1/3}}/3) \right) \right] ,  \,\ \Delta > 0 \, ,
\end{equation}
\begin{equation}
\label{phinrad}
\phi = (\alpha_{1o} \alpha_{2o} \alpha_{3o})^{-2/3} 
{\rm exp} \left[\frac{c_{3}}{\sqrt{-\Delta}} \, {\rm arctan} \, 
\left( \frac{1}{\sqrt{-\Delta}}
(c_{2} e^{2\eta} - 2\pi M_{ _{1/3}}/3) \right) \right] , \,\ \Delta  < 0 \, ,
\end{equation}
or
\begin{equation}
\label{phid0rad}
\phi = (\alpha_{1o} \alpha_{2o} \alpha_{3o})^{-2/3} 
{\rm exp} \left[\frac{- c_{3}}{c_{2} e^{2 \eta} -2 \pi M_{ _{1/3}}/3}  
\right] , \,\ \Delta = 0 \, .
\end{equation}
{}From Eq. (\ref{hiconsts}) one observes that if $\Delta \le 0$, then  
$3+2\omega \le 0$ in order to have real solutions
($h_{2}, \, h_{3}$ being real numbers) for the scale factors  
Eq. (\ref{ainrad}) or  Eq. (\ref{aid0rad}).  

Eqs.(\ref{airad}-\ref{phid0rad}), together with Eq. (\ref{rho}), 
show the most general solution for the Bianchi model V 
containing a radiation field or a fluid of relativistic 
particles.  For some particular values of our constants 
$c_{1},~c_{2},~c_{3}$, and $h_{i}$, they were firstly presented in 
Ref. \cite{Lo84c}.  The solution for $i=1$ is the most general solution 
for the FRW open Universe, cf. \cite{Lo84a}. In the particular case that 
$c_{1}=-c_{2}$ and $h_{i}=0$ one gets the open FRW solution given in 
Ref. \cite{Ba93}, where a misprint is found: 
$\beta \rightarrow -\beta$ in Eq. (76) of that paper.

In the particular case that $c_{1}=c_{2}=0$ one obtains the 
isotropic solution with $\phi<0$.  In this case, one has 
that $H(\eta)=$const., $\psi(\eta)=$const. and since 
$a(\eta) d\eta=dt$, one has that $H_{1}(t)=H_{2}(t)=H_{3}(t)=1/t$ 
with $a_{1}=a_{1o} t, ~a_{2}=a_{2o} t, ~a_{3}=a_{3o} t$, 
getting that $\phi=-\frac{4 \pi M_{ _{1/3}}}{3 a_{1o}^{2}t^{2}}$, a 
sign of antigravity?, a particular solution found 
elsewhere \cite{ObCh70s}.

For if $c_{3}=0$, the solution becomes the same as in the GR 
theory with $\phi=$const$\equiv M_{Pl}^{2}$,  
first obtained in Ref. \cite{Ru77}.  Further, if $h_{i}=c_{3}=0$, 
one obtains the $k=-1$ isotropic solutions of GR discussed in the 
late 60's \cite{grfrwk-1}.

\section{SOLUTIONS ANALYSIS \label{solana}}

The solutions presented in the preceding section show various 
behaviours depending upon the value of the BD parameter 
($\omega$), the type of fluid present ($\nu$), the amount of 
matter ($m_{\nu}$), and the anisotropic degree ($h_{i}$) of the models; 
let us consider each of 
them.  In the BD theory measurements imply that $\omega>500$, 
cf. Ref. \cite{Wi93}.  However, in general scalar tensor theories 
the coupling constant can be much less than 
one \cite{ig1,ig2}.  If such a theory is to be valid during some 
epoch in the early Universe, then it can be mimicked by an 
effective BD theory, but with its proper value of 
$\omega$.  Therefore, it is interesting to consider both great
and small $\omega$-values.  

In the preceding section we have presented two types of solutions 
for models with $\nu\neq 1/3$ and $\nu = 1/3$.  Following, 
we present the analysis of these solutions separately. 

\subsection{The case $\nu = 0$ \label{no1/3}}

Within the most interesting barotropic equations of state ($p=\nu \rho$) with 
$\nu\neq 1/3$ are the cases of dust ($\nu=0$), stiff fluid or supernuclear 
density ($\nu=1$) \cite{Ze6270}, particle creation ($\nu=2/3$) \cite{ObPi78}, 
and vacuum energy ($\nu=-1$).  For concreteness we shall consider in this 
subsection a gas of nonrelativistic massive bosons that could have played a 
dominant role (in the stress energy tensor) for the dynamics of the very early
universe.  This gas behaves as dust with an equation of state determined by 
$\nu=0$ \cite{Tu83}.  Accordingly, the total mass of the gas is given by the 
quantity $M_{o}=\frac{(3+2\omega)}{8 \pi} m_{o}=\rho_{o} a^{3}_{o}$.  

The values of $h_{i}$ are related to the degree of anisotropy present 
in the model.  As pointed out already, if $h_{i}=0$ then  $\Delta=0$  
implying that $a_{1}=a_{2}=a_{3}$, to have the open FRW 
model.   Anisotropic solutions are found for the case that 
the above constants fulfill that $\Delta\neq 0$.   For if $\Delta>0$ with
$\nu=0$, then $\omega<0$, cf. Eq. (\ref{dis5}).  In this case the  
solution is given by Eqs. (\ref{a15}, \ref{a25}, \ref{a2a3}) implying that
it is valid only in the interval\footnote{Note that for $\nu=0$ one has that 
$\eta = {\rm const.} ~t$. \label{f5}}
$\frac{-\sqrt{-2/\omega}~|h_{2}| - \eta_{o}}{m_o} < \eta <  
\frac{\sqrt{-2/\omega}~|h_{2}| - \eta_{o}}{m_o}$.   Then, for this solution to 
evolve long times one should demand that 
$\frac{\sqrt{2} \, |h_{2}|}{8 \pi M_{o}} \gg \frac{\sqrt{-\omega}}{|3+2\omega|}$, 
that is, this would be favored by large anisotropies ($h_{2}$), small masses, or
small negative $\omega$-values.  This 
solution is for $h_{2}>0$ deflationary in the scale factor $a_{2}$ and 
inflationary in $a_{3}$; for $h_{2}<0$ the roles of $a_{2}$ and 
$a_{3}$ are inverted.  However, there is no exit of inflation because
of the continuously increasing $arctanh$ function in Eq. (\ref{a25}).  
Therefore, this behaviour can be only useful if after some inflation 
time some other physical source (e.g. another scalar field) begins to
dominate the dynamics to stop the otherwise eternal inflationary 
behaviour. 

The most interesting case seems to be for $\Delta < 0$, which is 
actually the case for $\omega>0$.  The solution is given by Eqs. 
(\ref{a15}, \ref{a2n5}, \ref{a2a3}) and is valid in the whole 
interval${}^{\ref{f5}}$ $-\infty < \eta < \infty$ without restrictions.  For 
$h_{2}>0$ this solution is initially inflationary 
in $a_{2}$ and deflationary in $a_{3}$; again their roles can be 
inverted by choosing $h_{2}<0$.  In contrast with the $\Delta>0$ solution, in 
the present case the solution has exit, since the $arctan$ function in 
Eq. (\ref{a2n5}) tends asymptotically to $\frac{\pi}{2}$.  Thus, during some 
time interval this function behaves inflationary, after which a FRW behaviour 
follows.  

The strong exponential expansion is possible, though no cosmological 
constant or function is present; this is because the non-minimal coupling
in Eq. (\ref{bd}) implies new kinetic terms to the dynamical equations in 
comparison with GR \cite{Le95}.   In order to achieve a successful 
inflationary scenario one should demand 
that the three scale factors, or its volume,  inflate simultaneously.   This 
is apparently not possible from the form of the solution
for $a_{1}$, Eq. (\ref{a15}), which is a power-law standard expansion type.  
An inflationary stage may occur, however,  when the denominator of 
Eq. (\ref{hi}) is effectively a constant ($\psi \approx$ const.), while the 
numerator is a linear function, so $H_{i} \sim \eta$.  This 
behaviour can be attained only during some time interval, since $\psi$, 
given by Eq. (\ref{psisol}), is a quadratic function and is 
eventually numerically larger than the numerator 
to give rise to $H_{i} \sim 1/\eta$.   To get enough e-foldings 
($\frac{a_{f}}{a_{o}} = e^{N}$) of inflation to solve the horizon and 
flatness problems,  one can try to adjust the model parameters 
$\omega$, $h_i$, and $m_{o}$.   Let us impose the necessary 
conditions for it.  On the one hand, 
$\psi \approx$ const. from $\eta=0$ until the time $\eta_{*}$, with  
$\eta_{*} \equiv \frac{2^{1/2}}{m_{0}} \left( \eta_{0}^{2}+
\frac{h_{2}^{2}}{\omega}\right)^{1/2} - \frac{\eta_{0}}{m_{0}}$. On the
other hand, to last sufficient amount of e-foldings of 
inflation\footnote{For initial conditions with lower energy scales than 
Planckian, the number of e-folds can be smaller than 
$68$ \cite{KoTu90}.  In our models we pursue to identify the asymptotic
value of the scalar field with the Newtonian constant $G$. Therefore, the
energy scale $G^{1/2}$ appears.} ($N\sim 68$) the accelerated 
expansion must run at least until the time $\eta_{f}$, with 
$\eta_{f} \equiv  \frac{1}{m_{0}} \left[(\eta_{0}-h_{i})^{2} +2 N (\eta_{0}^{2}+
\frac{2 h_{2}^{2}}{\omega})\right]^{1/2} - \frac{\eta_{0}-h_{i}}{m_{0}}$.
Thus, the inflationary epoch is valid in the interval $0\le \eta < \eta_{f}$, 
provided that $\eta_{f} < \eta_{*}$.  Unfortunately, the latter inequality
is valid for very restricted values of $\omega$, in fact for
$\omega < - \frac{2 h_{2}^{2}}{\eta_{0}^{2}}$.  Otherwise, for arbitrary $\omega$ 
the number of e-foldings is at most $N=1/2$ only.   For instance, one may 
try to fit the parameter $m_o$ (if $m_o$ is augmented, $N$ grows at a fix time) 
to get the desired number of e-foldings, but then the above inequality is no
more valid. Analogously, whereas arbitrary large  values of 
$\frac{h_{2}^{2}}{\omega}$ can make arbitrary large the inflation 
time \cite{CeCh99}, the time $\eta_{f}$ grows in the same proportion, avoiding 
to have $\eta_{f} < \eta_{*}$.  In BD FRW cosmologies a similar result is found, 
where only negative values for $\omega$ can accomplish an accelerated, 
successful expansion \cite{Le95}: A solution to the flatness problem 
is not possible because the current value $\omega=500$ \cite{Wi93} resctricts 
the solution types, not allowing an accelerated expansion,  and not allowing 
to have nowadays ($n$) a value of $\phi_{n}^{-1}= 16\pi G$, $G$ being the 
Newtonian constant  \cite{LeFr93-4}.

\begin{figure}
\unitlength 1in
\centering
\begin{picture}(+6.5,6.5)(0,1)
\epsfxsize=5.0in \epsfbox{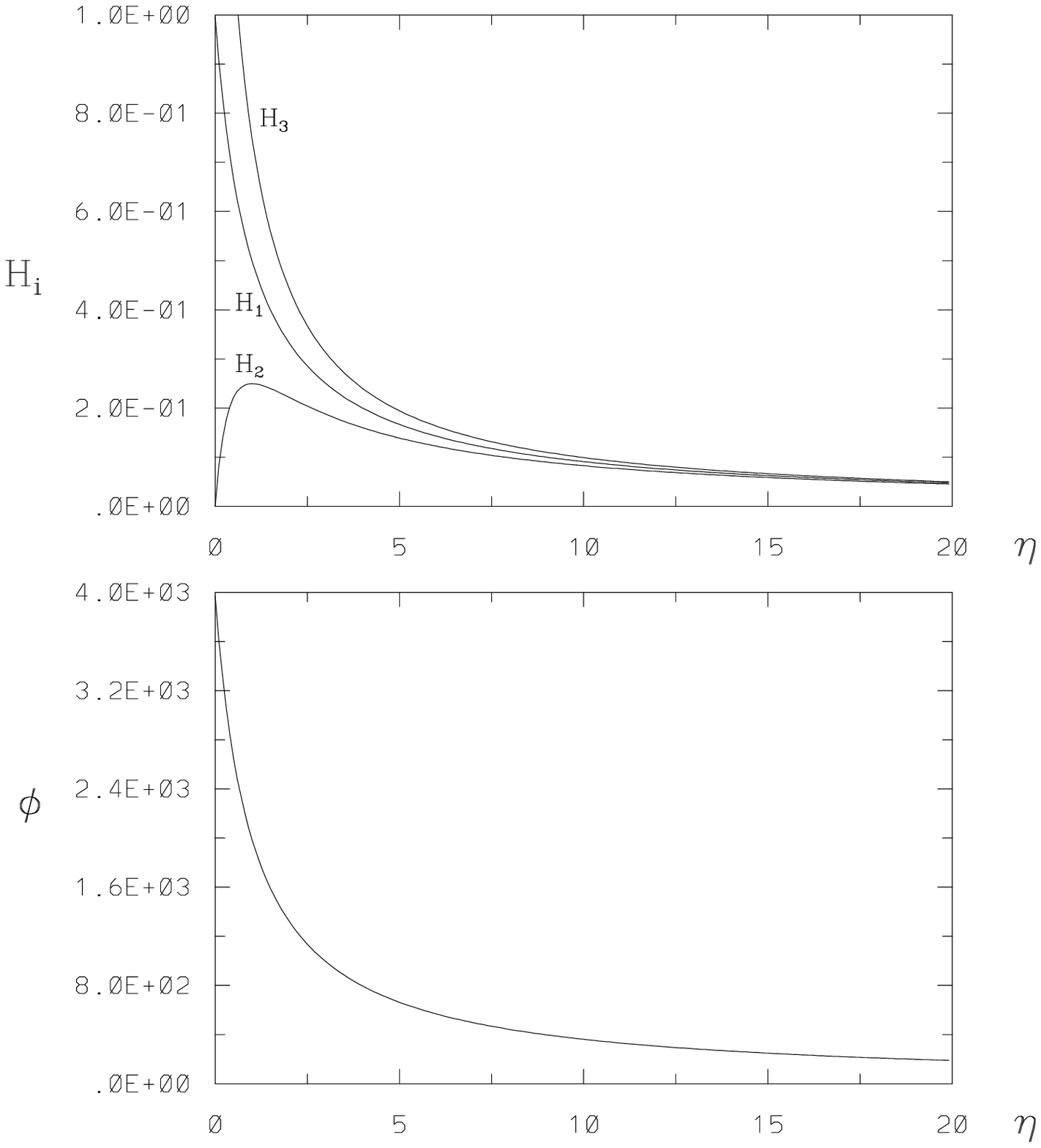}
\end{picture}
\vskip 1.5cm
\caption{
\label{fig1} }
\end{figure}

\begin{figure}
\unitlength 1in
\begin{picture}(6.5,6.5)(0,1)
\epsfxsize=5in \epsfbox{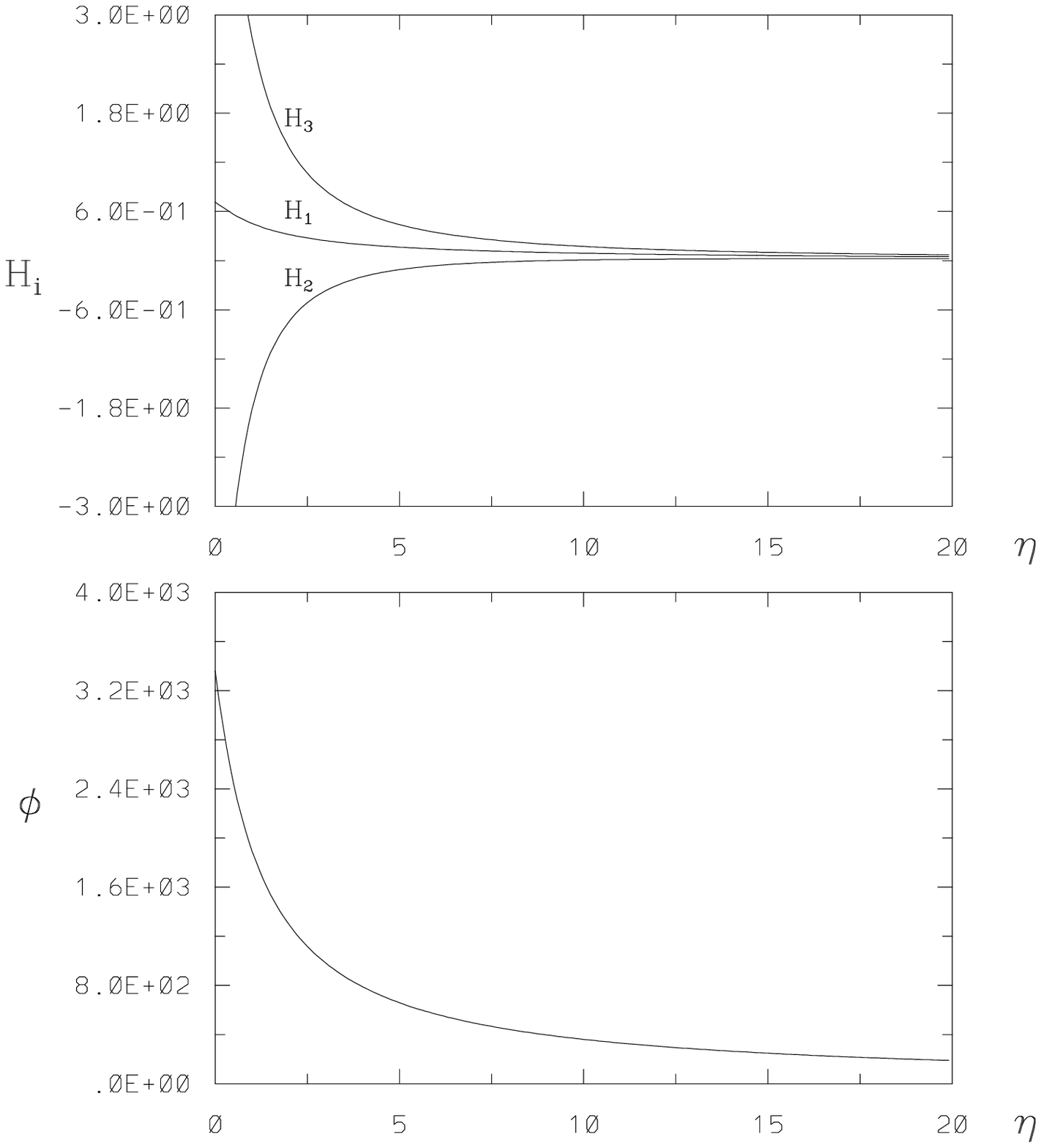}
\end{picture}
\vskip 1.5cm
\caption{
\label{fig2}}
\end{figure}

\begin{figure}
\unitlength 1in
\begin{picture}(6.5,6.5)(0,1)
\epsfxsize=5in \epsfbox{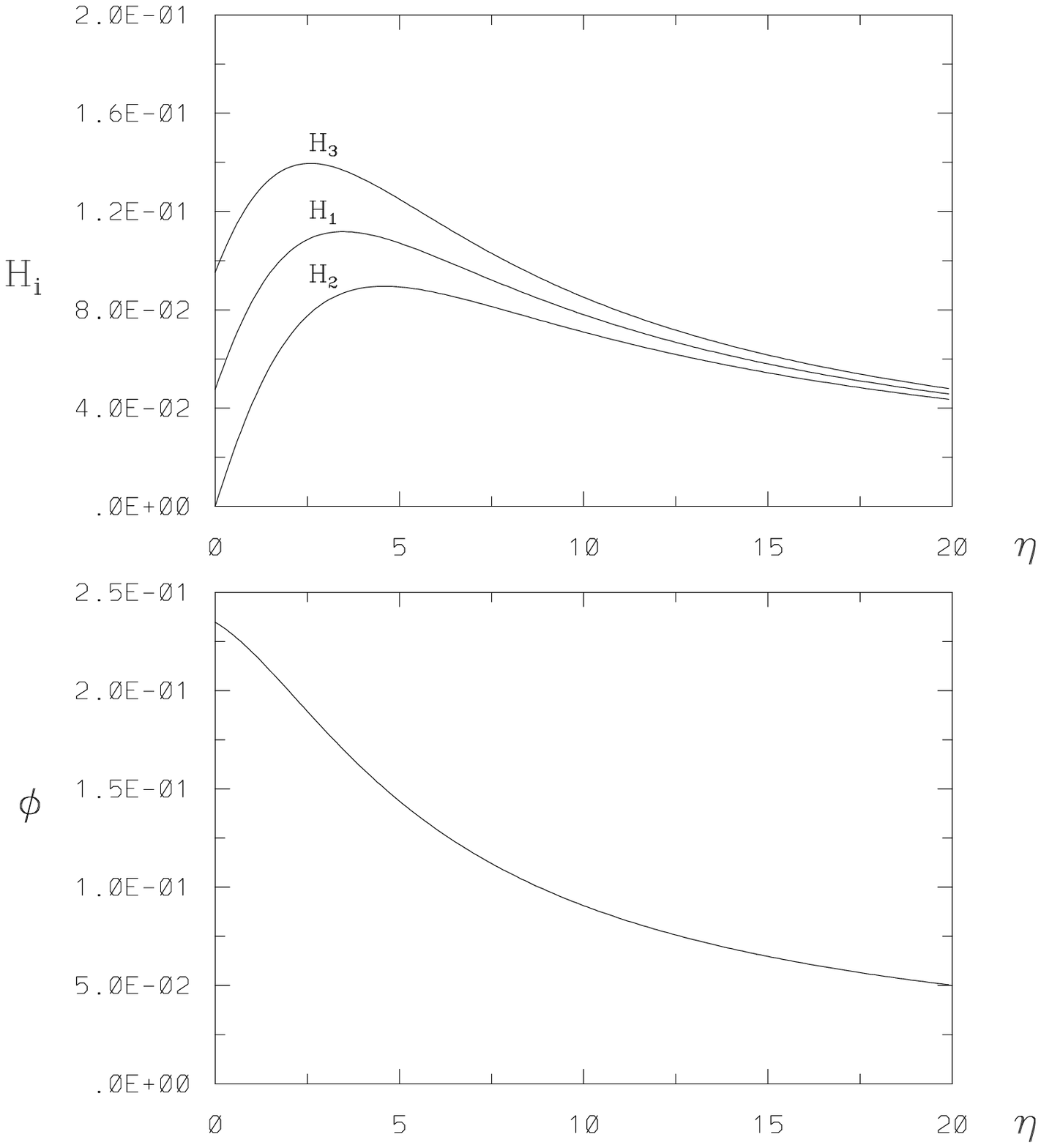}
\end{picture}
\vskip 1.5cm
\caption{
\label{fig3}}
\end{figure}

In figures \ref{fig1}-\ref{fig3} analytic solutions are plotted for some 
cosmological parameters.  Figure \ref{fig1} shows the Hubble
parameters and the BD field for $\nu=0$, $\omega=500$, and
$h_{2}=\eta_{o}=m_{0}=1$.  Initially the Hubble parameter 
$H_{2}$ grows almost linearly until $\eta = 0.42$ and, therefore, the 
solution is inflationary in this direction with 
$a_{2}\sim e^{{\rm \kappa_1} \eta + {\rm \kappa_2} \eta^{2}}$, where 
$\kappa_1, ~\kappa_2$ are constants.  
Afterwards $H_{2}$ continues growing at a lower rate until it 
stops at $\eta=1.0$ .  After this time this Hubble parameter is dominated 
by its denominator, cf. Eq. (\ref{hi}), and the solution becomes   
$H_{2}\sim\frac{1}{\eta}$.  Later on, after $\eta=1.6$ all three
$H_{i}$ are dominated by its denominator and the solution is of FRW type.

Figure \ref{fig2} shows the solution for $\omega=500$, 
$\nu=0$, $\eta_{o}=m_{0}=1$, and $h_{2}=10$. In this case, the initial 
anisotropy parameter was augmented, provoking an initial contraction of 
$a_{2}$ ($H_{2o}<0$), after which an expansion follows.  If one leaves above 
parameters fixed, but one augments the value of $m_o$, $a_2$ goes from 
a contraction to an inflation period.  

Finally, figure \ref{fig3} is similar to figure \ref{fig1}, but in 
figure \ref{fig3} a small $\omega$ was chosen to mimic a model 
of induced gravity valid when its potential plays no role, 
see Ref. \cite{CeCh99}.  In figure \ref{fig3} we observe an initially 
inflationary behaviour for the three scale factors, its exit, and 
its evolution to the open FRW.  Again the same effect is attained by 
augmenting $m_o$, the three scale factors inflate but with a restricted  
number of $N$ e-foldings, as mentioned above.  

Note that during the inflationary era the solutions 
are still anisotropic; they tend to isotropize once 
$H_{i}\sim 1/\eta$, see figures \ref{fig1}, \ref{fig2}, 
and \ref{fig3}.  

For small $\omega$'s, like in some induced gravity 
models\footnote{Our solutions are applicable in the induced gravity theory 
during a time interval, when its potential is not significant for the 
dynamics \cite{CeCh99}.} \cite{ig1},  the three Hubble parameters grow, 
as can be observed from figure \ref{fig3}.  For $\omega \ll 1$, 
cf. Ref. \cite{ig2}, the models isotropize during the inflationary 
era \cite{CeCh99}; later on, they evolve asymptotically to the open 
FRW model.

The dimensionless shear parameter, Eq. (\ref{dlsig}), turns out to be
\begin{equation}
\label{dlsigo}
\frac{\sigma}{H^{2}} = - \frac{6 h_{2}^{2}}{[m_{\nu} \eta -\eta_{o}]^{2}}
\,\  ,
\end{equation}
which for the $\Delta < 0$ solution,  one observes 
an asymptotic, isotropic behaviour $\sigma/H^{2} \to 0$, when 
$\eta \to \infty$, to yield the following solution: 
\begin{eqnarray}
&& a_{1} \rightarrow \sqrt{\frac{2}{2+\omega}} \,\ \eta \quad , \nonumber \\
&& a_2 \rightarrow e^{-\sqrt{\frac{\omega}{2}} \, \frac{\pi}{2}} \, a_1 \quad ,
\nonumber\\
&& a_3 \rightarrow e^{ \sqrt{\frac{\omega}{2}} \, 
\frac{\pi}{2}} \, a_1 \quad , \nonumber\\
&& \phi \rightarrow \left(\frac{2+\omega}{2}\right)^{\frac{3}{2}} \, 
\frac{m_{o}}{\eta} \quad ,
\end{eqnarray}
the bigger $\omega$ is, the larger (smaller) is the ratio between $a_{1}$ and  
$a_{2}$ ($a_{3}$).   In our graphics we observe only small 
differences in the $H_{i}$ ($i=1,2,3$) for $\eta\sim 20$ to get, in the 
asymptotic limit, the solutions found in Refs. \cite{DeOb70s,Le74,ChGu86}.  
This asymptotic solution is different from the open FRW solution 
in GR, and is described by a fixed point in the autonomous phase 
plane ($\dot{a}/a, \dot{\phi}/\phi$) analysis for FRW cosmologies in 
the BD theory \cite{HoWa98}.

\subsection{The case $\nu = 1/3$ \label{si1/3}}

The general solution found for $\nu=1/3$ is valid for a fluid of 
ultra-relativistic particles and/or radiation.   The isotropic, open solutions
are obtained when $h_{i}=0$, and for some particular values of our constants
they are reported in Refs. \cite{Lo84a,Ba93}.   Solutions given by 
Eqs. (\ref{airad})-(\ref{aid0rad}) have a similar structure: they are 
multiplied  by the power-law term $\psi^{1/2}$, as well as by the exponential 
term.  Initially the exponential term maybe more significant, and from it one 
could possibly obtain an inflationary stage in the evolution.   The time 
asymptotic behaviour is 
dominated by the power-law term.  The details  depend on the value 
of the discriminant ($\Delta$).   For the $\Delta = 0$ case the number of 
e-foldings of inflation is given by 
\begin{equation}
\label{N-rad-d0}
N = \frac{c_{3}}{2 c_{2} e^{2 \eta_{f}} - 
\frac{4 \pi M_{ _{1/3}}}{3}}- 
\frac{c_{3}}{2 c_{2}  - \frac{4 \pi M_{ _{1/3}}}{3}}   + 
\frac{1}{2} \, {\rm ln} \frac{\psi_f}{\psi_o}
\end{equation}
where $\eta_{f}$ is the time when the power-law solution begins to dominate, i.e., 
when the exponential term tends to one.  The number of e-foldings depends
finely on the value of the constants $c_{2}$,  $c_{3}$, and $M_{ _{1/3}}$, subject 
to the following inequalities: $(i)$ $c_{2} > 2 \pi M_{ _{1/3}}/3 > 0$,  the first inequality
is to avoid the point where $a_{i}$ becomes infinity, and the second is to have
$a_{i}, \rho >0$.  Note that $c_{1}>0$ since 
$c_{1} c_{2} = \left(2 \pi M_{ _{1/3}}/3\right)^{2}>0$; $(ii)$ 
$c_{1} + c_{2} > 4 \pi M_{ _{1/3}}/3$ to have $\psi_{o}, ~a_{io} >0$; and
$(iii)$ $c_{3} <0$ to have $N>0$.  The right choice of these parameters gives
the desired number of e-folding of inflation ($N$).   Unfortunately, the present
case ($\Delta=0$) is valid for $\omega\le-3/2$, otherwise $a_{2}$ and $a_{3}$ 
become complex functions, cf. Eqs. (\ref{hiconsts}) and (\ref{aid0rad}).   Recall that values for  $\omega<-3/2$ correspond to a negative scalar field density in the conformally rescaled Einstein frame 
($\hat{g}_{\mu \nu} = G \phi g_{\mu \nu}$), see for 
instance \cite{HoWa98}.  However, if one believes that the correct 
physics is in the Jordan frame, negative 
$\omega$'s maybe still interesting.  Recall that changing frames via 
conformal transformations imply physically different gravity theories since 
the matter couplings are different, as well as geodesic equations, see 
for instance \cite{So95}. 

Solutions with $\Delta>0$ have the same problem as in the 
$\nu\neq\frac{1}{3}$ case: they inflate anisotropically during the whole
evolution given by the time interval

${\rm ln} \left[ \left(2 \pi M_{ _{1/3}}/3 -1 \right) 
\left( \frac{2 \pi M_{ _{1/3}}}{3c_{2}^{2}} - 
\frac{c_{1}}{c_{2}}\right)^{1/2} \right]^{1/2} 
< \eta < 
{\rm ln} \left[ \left(2 \pi M_{ _{1/3}}/3 +1 \right) 
\left( \frac{2 \pi M_{ _{1/3}}}{3c_{2}^{2}} - 
\frac{c_{1}}{c_{2}}\right)^{1/2} \right]^{1/2} $
and the graceful exit problem remains because inflation never ends. 

\begin{figure}
\unitlength 1in
\begin{picture}(6.5,6.5)(0,1)
\epsfxsize=5in \epsfbox{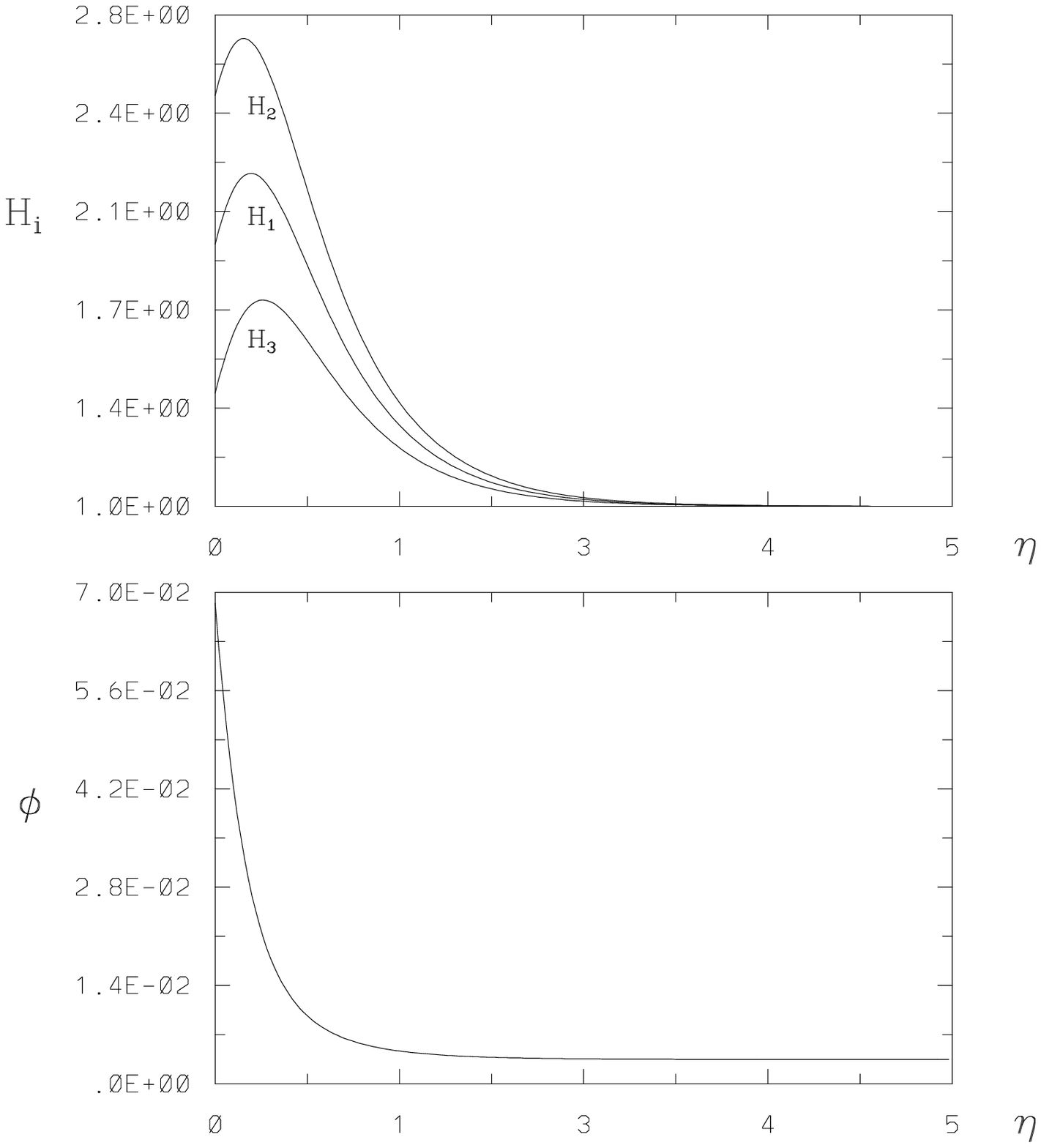}
\end{picture}
\vskip 1.5cm
\caption{
\label{fig4} }
\end{figure}

\begin{figure}
\unitlength 1in
\begin{picture}(6.5,6.5)(0,1)
\epsfxsize=5in \epsfbox{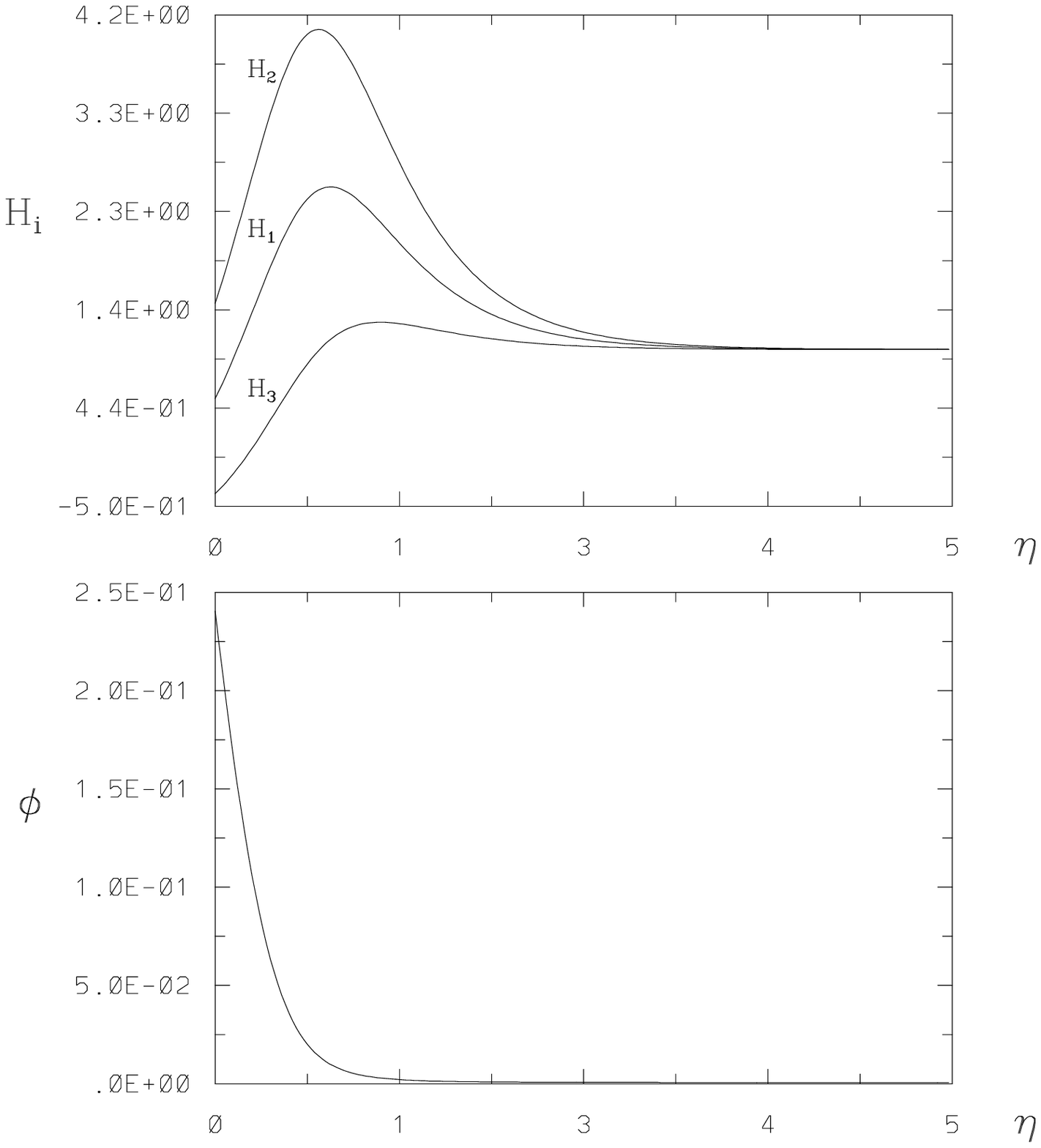}
\end{picture}
\vskip 1.5cm
\caption{
\label{fig5} }
\end{figure}

Solutions with $\Delta<0$ imply that $\omega<-3/2$ in order to have 
real (instead of complex) solutions for the scale factors in accordance with 
Eqs. (\ref{hiconsts}) and (\ref{ainrad}).   Figures \ref{fig4} and  \ref{fig5} show
solutions for the different values of the physical parameters.   One 
notes that terms containing  $c_{2} e^{2 \eta}$ dominate the dynamics 
very rapid, and therefore shall account for the major contribution to the 
dynamics in asymptotic times.  Terms with $c_{1} e^{-2 \eta}$  could   be 
important for the very beginning.  In figure \ref{fig4} we have chosen the parameters to be $\omega=-2$, $h_{2}=c_{1}=c_{2}=1$, and $m_{ _{1/3}}=-1$, 
which corresponds to $M_{ _{1/3}}=+1/(8 \pi)$ to have a positive density.  
In figure \ref{fig5} we have augmented the value of the anisotropic 
parameter $h_{2}$ (=100), and the 
values of $c_{1}$ and $c_{2}$ ($c_{1}>c_{2}$) as well, to observe the 
influence of the $c_{1} e^{-2 \eta}$ term. Otherwise, if $c_{2}>c_{1}$ the 
solution becomes the isotropic solution yet from the very beginning. 

\begin{figure}
\unitlength 1in
\begin{picture}(2.5,5.5)(-1,0)
\epsfxsize=4in \epsfbox{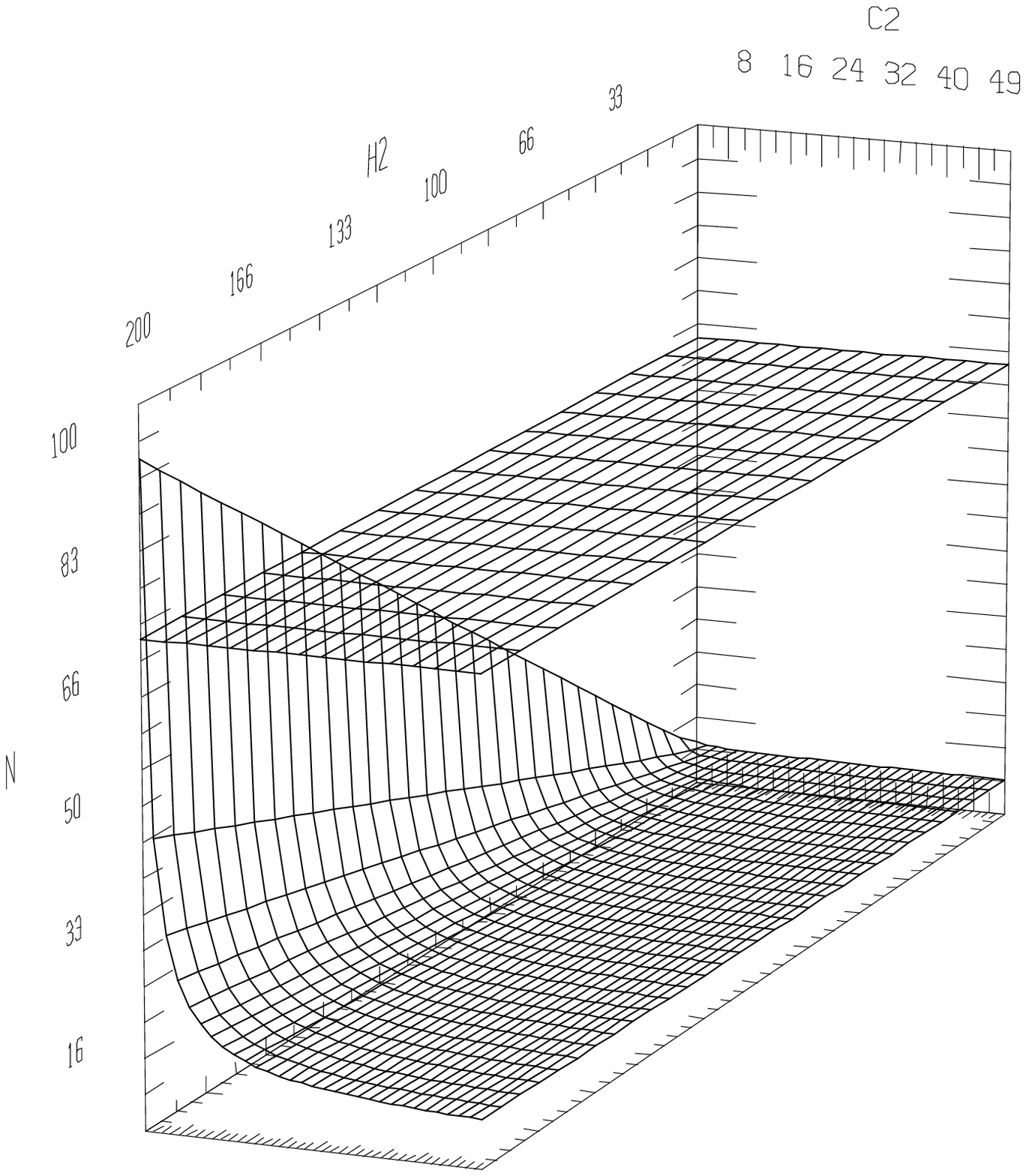}
\end{picture}
\vskip -1.0cm
\caption{
\label{fig6} }
\end{figure}

The solutions with $\Delta<0$ have the chance to achieve a number of e-foldings 
of inflation ($\frac{a_{f}}{a_{o}} = e^{N}$) given by
\begin{equation}
\label{N-rad-d<0}
N = \left[\frac{-c_{3}}{2 \sqrt{-\Delta}}\right]  
\left[ 
{\rm arctan} \frac{1}{\sqrt{-\Delta}} \left(c_{2} e^{2 \eta_{f}} - 
\frac{2 \pi M_{ _{1/3}}}{3}\right) - 
{\rm arctan} \frac{1}{\sqrt{-\Delta}} \left(c_{2}  - 
\frac{2 \pi M_{ _{1/3}}}{3} \right)\right] + 
\frac{1}{2} \, {\rm ln} \frac{\psi_f}{\psi_o}  \, .
\end{equation}
One can choose the integration constants to have the desired number of 
e-foldings, subject to following constrains 
(the arguments are the same as in the $\Delta=0$ case): 
$(i)$ $c_{2} >0$, $M_{ _{1/3}} >0$.  Note that $\Delta<0$ implies that 
$c_{1} c_{2}>\left(2 \pi M_{ _{1/3}}/3\right)^{2} > 0$, hence, $c_{1} > 0$; 
$(ii)$ $c_{1} + c_{2} > 4 \pi M_{ _{1/3}}/3$; and $(iii)$ $c_{3} <0$.   Accordingly, 
figure \ref{fig6} shows the parameter dependence of the number of e-foldings 
of inflation ($N$).  Small values of $c_{2}$ together with big $h_{i}$ values are 
requested to achieve a sufficient number for N.

The dimensionless shear parameter, Eq. (\ref{dlsig}), turns out to be
\begin{equation}
\label{dlsig13}
\frac{\sigma}{H^{2}} = - \frac{6 h_{2}^{2}}
{[-c_{1} e^{- 2 \eta} + c_{2} e^{2 \eta} - c_{3}]^{2}}
\,\  ,
\end{equation}
which for the $\Delta \le 0$ solutions, their asymptotic behaviour, 
when $\eta \to \infty$, implies $\sigma/H^{2} \to 0$,  yielding the 
following isotropic solution: 
\begin{eqnarray}
\label{hap}
&&H_{i}(\eta)\rightarrow 1, \quad a_{i}(\eta) \rightarrow a_{io} \, 
\sqrt{c_{2}} \,  e^{\eta}, \quad \phi(\eta) \rightarrow 
(a_{1o}a_{2o}a_{3o})^{-2/3} \qquad \Rightarrow \nonumber \\
&&H_{i}(t) \rightarrow \frac{1}{t}, \quad a_{i}(t) \rightarrow 
a_{io}\, t , \quad ~~~~~~~
\phi(t) \rightarrow (a_{1o}a_{2o}a_{3o})^{-2/3} \, .
\end{eqnarray}
Independent of the values all parameters may have, the solution evolves to 
its isotropic behaviour \cite{ObCh70s}, as can be seen in figures \ref{fig4} and 
\ref{fig5}.  The above limit solution, Eq. (\ref{hap}), tends to the radiation 
solution in GR for asymptotic times.

\section{FLUCTUATIONS OF THE BD FIELD \label{fluc}}

Although we are working in a framewok of a classical theory, one may also 
consider quantum fluctuations of the BD field during inflation. It is 
well known that isocurvature and curvature fluctuations are generated, and
they produce perturbations in the density and in the CMBR.  Fluctuations
of the BD field has been analised \cite{KoSaTu90,Pbd97} mostly in the Einstein 
frame\footnote{The transformation of the metric is 
$\hat{g}_{\mu \nu} = G \phi g_{\mu \nu}$ and the new field 
$\hat{\phi} = \sqrt{\frac{3+2\omega}{16 \pi G}} \, 
{\rm ln}(G \phi)$ \label{ef}}, 
since there the new scalar field ($\hat{\phi}$) is minimally 
coupled and its the kinetic term 
is canonical, therefore, the role of fluctuations is easy to interprete   
physically, and to import results from what is known in the Einstein 
theory \cite{SaBoBa89,KoSaTu90}.  For example, if it were the case that we
consider an additional effective constant potential (false vacuum energy) 
in the Jordan frame, this would turn out to behave like an exponential
potential in the Einstein frame, i.e., a la extended 
inflation \cite{KoSaTu90}.  In this case, one gets that 
$\delta\rho/\rho$ depends on $\lambda^{4/(2\omega-1)}$ due to 
power law inflation, where $\lambda$ is some perturbation scale 
\cite{Per85,KoSaTu90,Pbd97}.  Then, all machinery
of reconstructing the inflation potential could be 
applied \cite{Reconst98}.  In our case, one does not have a 
potential that forces $H=$ const., but in our models one has that 
$H \sim t$ during some initial stage.  Because of this fact, one expects 
that the spectrum is non-scale invariant too.  In our inflationary models, 
however, if one considers the Einstein frame to analise the 
perturbations, one finds that 
the transformation factors involve quantities with $\sqrt{3+2\omega}$ (see footnote \ref{ef}), which in our solutions with inflation, it is a complex number.  Then, one cannot correctly interprete our inflationary solutions 
in the Einstein frame.  Furthermore, values of $\omega<-3/2$ generate negative scalar-field densities in the conformally rescaled Einstein  frame \cite{HoWa98}. 

For $\omega>-3/2$ we do not have an inflationary behaviour, therefore, the
production of quantum fluctuactions is out of interest because the horizon
problem remains.  However, if one starts the anisotropic models with
Planckian initial conditions one can explain \cite{Ba95} the observed 
anisotropy of the CMBR measured by COBE \cite{Cobe94}.

\section{CONCLUSIONS \label{con}}

Our analysis is based on the exact solutions reported
in this work and in Ref. \cite{ChCe95}.  For a fluid with $\nu \neq 1/3$
the particular solution Eq. (\ref{psisol}) was used, but other solutions
may be found from Eq. (\ref{psi1}), which is considerable simpler than 
other equations found  in the past \cite{Lo84c}.  For instance, our time 
asymptotic solution does not tend to the GR one, therefore, the latter must 
correspond to another solution of Eq. (\ref{psi1}).  For the radiation case 
($\nu = 1/3$) the most general solution is given.  The FRW solution is same 
as the given by our $i=1$ case (recall that $h_{1}=0$).  The GR Bianchi V 
case is achieved when $c_{3}=0$. 

The above-presented exact solutions have a variety of possible behaviours
depending on the values of the physical parameters $\omega$, $\nu$,
$m_{\nu}$, and $h_{2}$.  Within  some parameter range  it is 
possible to encompass a cosmological model that incorporates the issues 
mentioned in the introduction: to have anisotropic initial conditions 
that leads to an inflationary stage, and, as time goes on, to
isotropize towards an open FRW model.  In this way, inflation
represents a transient attractor, and the FRW behaviour is an asymptotic 
attractor.   Further, these models can explain the observed anisotropy degree 
of the CMBR measured by COBE \cite{Cobe94}, if the model is to be started 
with Planckian initial conditions \cite{Ba95}.

It is peculiar that these solutions show an inflationary behaviour, even 
without a cosmological constant or function. This is because the 
non-minimal coupling in Eq. (\ref{bd}) implies new kinetic terms to the 
dynamical equations in comparison with GR \cite{Le95}.   For the 
non-radiating ($\nu \ne 1/3$),  anisotropic 
solutions with $\Delta \neq 0$ the inflationary stage takes place when
the denominator of Eq. (\ref{hi}) is almost a constant, whereas the numerator 
is linear.  This is achieved during some time interval depending on the 
physical parameters $\omega$, $\nu$, $h_i$, and $m_{\nu}$.  Especially, 
large values of $m_\nu$ or $\frac{h^{2}_{2}}{\omega}$ favor strong 
exponential expanding solutions of the type 
$a_{i} \sim e^{{\rm \kappa_1} \eta + {\rm \kappa_2} \eta^{2}}$, but the 
number of e-foldings of inflation cannot be greater than $1/2$ for 
$\omega>0$.    Solutions with $\Delta > 0$ are inflationary, but without 
exit, diverging at asymptotic times.  Solutions with $\Delta<0$,
independent of a possible inflationary behaviour, 
asymptotically isotropize to a FRW open model.   For $\omega \gg 1$ the 
isotropization mechanism happens after the (short) inflationary 
era (see figure \ref{fig1}), and for $\omega \ll 1$ it occurs during 
inflation;  some figures of the latter case are shown elsewhere \cite{CeCh99}.
For the radiating case ($\nu = 1/3$),  anisotropic solutions are found for
$\Delta \neq 0$ and $\Delta = 0$, as well.   Solutions with $\Delta > 0$ 
are inflationary again, but with the same properties as in the non-radiating
case: inflation is only present in one scale factor, and there is no exit
of it, i.e. the solution asymptotically diverges.  In the cases 
with  $\Delta \le 0$ inflation takes place, but again $\omega$ must 
be negative ($\omega \le -3/2$) to guarantee a sufficient amount of 
e-folds of expansion ($N\sim 68$).   The reason behind this can be found in 
Eqs. (\ref{gr}) and  (\ref{hig}).   For $\omega \le -3/2$ the sign of
some source terms (related to the trace $T$) and kinetic terms are 
reverse and play an inverse role than normally, making possible to have 
transient inflationary solutions.

It seems that the most interesting cosmological exact solutions in the BD 
theory are those with a BD parameter different from the desired 
$\omega>500$, in fact for negative $\omega's$.   For the latter case there 
have been reported even more solutions than with $\omega>0$, see for 
instance \cite{DeOb70s, ChCe95}.  However, values of $\omega<-3/2$ 
correspond to negative scalar field densities in the conformally rescaled 
Einstein  frame ($\hat{g}_{\mu \nu} = G \phi g_{\mu \nu}$) \cite{HoWa98}.  Therefore, such solutions have to be taken with caution.   Furthermore, 
within the BD theory alone is not possible to have 
reheating in the Universe, since
there is no potential nor any energy transfer to mass terms.  However, 
reheating is necessary after inflation, and therefore, a more general
theory than BD is in order.   In this context, we have analysed the case of 
an induced gravity theory, where after a mild inflation period (due to non-minimal coupling) the Bianchi V model isotropizes, afterwards inflation 
due to the potential of the theory follows and reheating takes place \cite{ig2,CeCh99}.  Our solutions may be also of physical interest in a scalar-tensor theory, perhaps some kind of scalar-tensor with $\omega(\phi)$, 
in which the solutions are to be valid during some (cosmic) time stage 
when the physical theory is mimiced by a BD effective action, yet with 
some other value for $\omega$, not restricted to be bigger than 500 at 
the begining of times.

Finally, we think that much work has to be done in the field of exact
solutions in the context of modern cosmology.  There are many solutions 
written in scaled variables, whose physics is hidden because of its 
complexity, and because less effort has been put on the analysis of its 
physical meaning.   Nice properties as inflation, graceful exit, and 
isotropization has been shown in the present work, and without any 
cosmological constant or function, but for a restricted range of physical 
parameters. 


\newpage

\begin{center}
{\bf Figure captions }
\end{center}
\bigskip\bigskip

Figure 1:  The upper graph shows the Hubble parameters as a function of 
the time $\eta$.  As the universe evolves the three Hubble rates tend 
to their corresponding open FRW solution $H_{1}$.  For these plots 
we have taken $\nu=0$, $\omega=500$, and
$h_{2}=\eta_{o}=m_{0}=1$.  For these values one has that $H_{1o} > 0$, 
$H_{3o} > 0$, and $H_{2o} = 0$.   In the figure below the evolution of the 
scalar field is shown.  For $\eta\gg 2.4$, $\phi \sim 1/\eta$  as in the FRW 
cosmology.  

\bigskip

Figure 2:  The upper graph shows the Hubble parameters as a function of 
the time $\eta$.  In these
plots we have taken the parameters to be the same as in figure \ref{fig1},
except for the anisotropy parameter that we have augmented to be 
$h_{2}=10$.  Because of this, $a_{2}$ undergoes a contraction that 
diminishes as time evolves.  At $\eta \approx 9$ the contraction stops 
and an expansion follows.  $a_{1}$ and $a_{3}$ always expand.   The 
figure below shows the scalar field as a function of the time $\eta$. 

\bigskip 

Figure 3: The upper graph shows the Hubble parameters as a function 
of the time $\eta$.  For these plots we have taken 
$\nu=0$, $h_{2}=\eta_{o}=m_{o}=1$, and $\omega=0.1$.  That is, we have 
changed $\omega$ to be a small number to show an example of the type 
of induced gravity without potential. The evolution of the scalar field is 
shown in the figure below.

\bigskip

Figure 4: The upper graph shows the Hubble parameters as a function 
of the time $\eta$ for the parameters 
$\omega=-2$, $h_{2}=c_{1}=c_{2}=-m_{_{1/3}}=1$.   For these 
values the three initial Hubble functions are positive, $H_{io}>0$.   The 
expansion begins super-inflationary until some turning point after which 
the three Hubble rates tend to the isotropic solution.    In the figure below 
the evolution of the scalar field is shown.  

\bigskip

Figure 5: The upper graph shows the Hubble parameters as a function of 
the time $\eta$ for the parameters 
$\omega=-2$, $h_{2}=c_{1}=100$, $c_{2}=10$, $m_{_{1/3}}=-1$.   For 
these values $H_{1o} > 0$,  $H_{2o} > 0$, and $H_{3o} < 0$.   $H_{1}$ and  
$H_{2}$ begin expanding, and $H_{3}$ contracting, until some turning point 
after which the three Hubble rates tend to the isotropic solution.    In the 
figure below the evolution of the scalar field is shown. 

\bigskip

Figure 6: The number of e-foldings ($N$) of exponential expansion as a 
function of $h_{2}$ and $c_{2}$.  This plot corresponds to an integration 
time from $\eta=0$ to $5$.  The plane $N=68$, representing the threshold 
of successful expansion, is shown for comparison. 

\end{document}